\documentclass[12pt]{article}

\usepackage[margin=1in]{geometry}
\usepackage[T1]{fontenc}
\usepackage[utf8]{inputenc}
\usepackage{lmodern}
\usepackage{microtype}

\usepackage{parskip} 

\usepackage{amsmath,amssymb,amsthm,mathtools,bm}
\usepackage{siunitx}
\usepackage{hyperref}
\hypersetup{
  colorlinks = true,
  urlcolor   = blue,
  linkcolor  = blue,
  citecolor  = blue
}

\usepackage{graphicx}
\usepackage{caption}
\usepackage{subcaption}
\usepackage{booktabs}
\usepackage{longtable}
\usepackage{tabularx}
\usepackage{array}
\usepackage{xcolor}
\newcolumntype{Y}{>{\raggedright\arraybackslash}X}
\usepackage{float}
\usepackage{threeparttable}
\usepackage{enumitem} 

\usepackage[numbers]{natbib}

\theoremstyle{plain}
\newtheorem{theorem}{Theorem}[section]

\theoremstyle{definition}
\newtheorem{definition}[theorem]{Definition}

\theoremstyle{remark}
\newtheorem{remark}[theorem]{Remark}

\newcommand{\R}{\mathbb{R}}
\newcommand{\RP}{\mathbb{RP}}
\newcommand{\SP}{\mathbb{S}}
\newcommand{\PGL}{\mathrm{PGL}}         

\newcommand{\GLp}{\mathrm{GL}^{+}}
\newcommand{\OPGL}{\mathrm{OPGL}}

\newcommand{\OPS}{\mathrm{OPS}}

\newcommand{\E}{\mathbb{E}}
\newcommand{\Var}{\mathrm{Var}}

\newcommand{\jdir}{j_{\mathrm{dir}}}
\newcommand{\jVW}{j_{\mathrm{VW}}}
\newcommand{\tSOPS}{tS_{n,\mathrm{OPS}}}
\newcommand{\tSPS}{tS_{n,\mathrm{PS}}}
\newcommand{\tSigmaOPS}{t\Sigma_{\mathrm{OPS}}}
\newcommand{\tSigmaPS}{t\Sigma_{\mathrm{PS}}}

\title{Extrinsic Total-Variance and Coplanarity via\\ Oriented and Classical Projective Shape Analysis}

\author{
  Musab Alamoudi\thanks{King Abdulaziz University, Jeddah,
  P.O.~Box~80200, Zip Code~21589, Kingdom of Saudi Arabia.}
  \and
  Robert L.\ Paige\thanks{Department of Mathematics and Statistics,
  Missouri University of Science and Technology, Rolla, MO 65409, U.S.A.}
  \and
  Vic Patrangenaru\thanks{Department of Statistics,
  Florida State University, Tallahassee, FL 32304, U.S.A, vic@stat.fsu.edu.}
}

\date{November 18, 2025}  

\begin{document}

\maketitle

\begin{abstract} Projective shape analysis provides a geometric framework for studying digital images acquired by pinhole digital cameras. In the classical projective shape (PS) method, landmark configurations are represented in $(\RP^2)^{k-4}$, where $k$ is the number of landmarks observed. This representation is invariant under the action of the full projective group on this space and is sign-blind, so opposite directions in $\R^{3}$ determine the same projective point and front--back orientation of a surface is not recorded. Oriented projective shape ($\OPS$) restores this information by working on a product of $k-4$ spheres $\SP^2$ instead of projective space and restricting attention to the orientation-preserving subgroup of projective transformations. In this paper we introduce an extrinsic total-variance index for OPS, resulting in the extrinsic Fr\'echet framework for the m dimensional case  from the inclusion $\jdir:(\SP^m)^q\hookrightarrow(\R^{m+1})^q,q=k-m-2$. In the planar pentad case ($m=2$, $q=1$) the sample total extrinsic variance has a closed form in terms of the mean of a random sample of size $n$ of oriented projective coordinates in $S^2$. As an illustration, using an oriented projective frame, we analyze the Sope Creek stone data set, a benchmark and nearly planar example with $41$ images and $5$ landmarks. Using a delta-method applied to a large sample and a generalized Slutsky theorem argument, for an OPS leave-two-out diagnostic, one identifies coplanarity at the $5\%$ level, confirming the concentrated data coplanarity PS result in Patrangenaru(2001)\cite{Patrangenaru2001}. \end{abstract}

\clearpage
\clearpage

\pagenumbering{arabic}
\setcounter{page}{1}
\section{Introduction}

Landmarked images obtained from pinhole cameras or similar optical systems can be studied using projective geometry. Three-dimensional scenes are mapped to the image plane by a central projection, and projective shape analysis studies the resulting landmark configurations modulo projective transformations ( see \citet{PaEl:2015} and references therein ). In the classical projective shape (PS) model, configurations are represented in $(\RP^m)^q$ and are invariant under the full projective group $\PGL(m{+}1,\R)$.

A basic feature of this model is that it is insensitive to the sign of homogeneous coordinates: the projective point $[x]\in\RP^m$ identifying the real and virtual directions $x$ and $-x$ in $\R^{m+1}$. For many applications this is appropriate, but in situations where a physical surface has a natural orientation, such as a rock face, a biological surface, or a road scene, the difference between front and back is meaningful. In such settings, oriented projective geometry and oriented projective shape ($\OPS$) provide a more faithful description \citep{MaPa:2005,ChPaPa:2022}.

The $\OPS$ framework replaces axes by directions on the unit sphere $\SP^m$ and restricts attention to orientation-preserving transformations $\OPGL(m)$. Choi et al(2022)\citep{ChPaPa:2022} defined OPS shape spaces, and represented them as products $(\SP^m)^q$ of spheres, thus allowing for a extrinsic or intrinsic Fr\'echet analysis on these spaces, linking OPS analysis with multivariate directional statistics.

The present paper focuses on \emph{extrinsic total variance} in the OPS case, aiming to measure and test for the overall dispersion of oriented projective coordinates. The motivating hypothesis is coplanarity: if a landmarked surface is flat, then after OP coordinate representation relative to a fixed oriented projective frame from the landmark configuration, the OP coordinates of the remaining landmarks configuration should be constant, thus their OPS total variance should vanish.

We extend work in the extrinsic Fr\'echet framework in \citet{PaEl:2015}, using the inclusion
\[
\jdir:(\SP^m)^q\hookrightarrow(\R^{m+1})^q
\]
and the usual Euclidean distance in the ambient numerical space. In the planar single- OP coordinate case ($m=2$, $q=1$), the resulting OPS total-sample variance index turns out to have a particularly simple form:
\[
\tSOPS = 2(1-R_n),
\qquad
R_n = \|\bar{u}\|,
\]
where $\bar{u}$ is the sample mean of the OPS unit vectors representing the oriented OP coordinates of the sample of size n of $k$-ads . This expression mirrors the Veronese--Whitney (VW) PS total sample variance $2(1-\lambda_1^{(n)})$ based on the largest eigenvalue of the mean of the empirical in the space of symmetric matrices (see Fisher et.al.(1996)\cite{FiHaJiWo:1996}).

Our approach is as follows: (i) we define the population OPS total-variance $\tSigmaOPS$ and its sample counterpart $\tSOPS$ using an extrinsic Fr\'echet framework, (ii) using a delta-method we derive the standard error and large-sample confidence intervals for $\tSigmaOPS$; (iii) we find confidence intervals for $\tSigmaOPS$, using both the chi-square and normal-approximations for the distribution of $\tSOPS$; (iv) using large sample theory, we tests for the coplanarity of the Sope Creek landmark data via the hypothesis $H_0:\tSigmaOPS=0$. The results confirm and extend those in \citet{Patrangenaru2001}, that were restricted to a small support of data assumption.

Section~\ref{sec:shape} reviews projective and oriented projective shape. Section~\ref{sec:ops} introduces the OPS extrinsic total-variance index and its asymptotic variance. Here one recalls the VW PS total variance formulas. Section~\ref{sec:test} formulates the OPS coplanarity test and leave-two-out diagnostic and presents the Sope Creek stone analysis. A short discussion appears in Section~\ref{sec:discussion}.

\section{Projective and Oriented Projective Shape}
\label{sec:shape}

In this section, following Mardia and Patrangenaru(2005)\cite{MaPa:2005} and Choi et al(2022)\cite{ChPaPa:2022} we briefly recall definitions of projective shape and oriented projective shape. We focus on the planar case $m=2$ and a single remaining landmark, in view of our application.

\subsection{Projective shape and its representation via projective frames}

Points in a numerical space are here presented as column vectors. Given $k\ge m+2$ , a $k$-ad in $\R^m$ is an ordered set of labeled points $(x_1, \dots, x_k) $ in $\R^m$. To each point $x\in\R^m$, we associate the projective point $[\tilde x]\in \mathbb R P^m,$ where
\[
\tilde x = (x^T, 1)^T \in \R^{m+1}
\]

A {\em{projective frame}}(or
{\it projective basis}) in an $m$ dimensional real projective space is an ordered set of $m + 2$ projective points in
general position. An example of projective frame is the {\it standard projective frame} is $([e_1],\dots,[e_{m +
1}],[e_1 + \dots + e_{m + 1}])$ in $\mathbb{R}P^m.$

In projective shape analysis it is preferable to employ coordinates
invariant with respect to the group PGL$(m+1)$ of projective
transformations of $\mathbb R P^m$. Recall that we define an element of $\beta=\beta_A \in PGL(m+1)$ given in terms of a matrix $A\in GL(m+1)$, defined
by its action on $\mathbb{R}P^m$ as follows: $\beta([x])=[Ax].$ The  projective transformation takes a projective
frame to a projective frame, and its action on $\mathbb R P^m$ is
determined by its action on a projective frame, therefore we
define the {\em{projective coordinate(s)}} of a point $p \in
\mathbb{R}P^m$ w.r.t. a projective frame  $\pi = (p_1,\dots, p_{m +
2})$ as being given by
\begin{equation} \label{projcoord}
p^{\pi} = \beta^{-1} (p), \end{equation} where $ \beta \in PGL(m+1)$
is the unique projective transformation taking the standard projective frame
to $\pi.$ These coordinates automatically have the projective invariance
property.
\begin{remark}\label{r:proj-coord} Assume $u, u_1, \dots, u_{m+2}$
are points in $\mathbb{R}^{m},$ such that $\pi = ( [\tilde{u}_1],
\dots, [\tilde{u}_{m+2}])$ is a projective frame. If we consider the
$(m+1) \times (m+1) $ matrix $U_m = [\tilde{u}_1^T, \dots,
\tilde{u}_{m+1}^T],$ the projective coordinates of $ p =[\tilde{u}]$
w.r.t. $\pi$ are given by
\begin{equation} \label{projcoord_formula}
 p^{\pi} = [ y^1(u): \dots: y^{m+1}(u)],
\end{equation}
where
\begin{eqnarray} \label{v}
v(u) = U_m^{-1} \tilde{u}^T
\end{eqnarray}
and
\begin{equation} \label{y}
y^j(u)=\frac{v^j(u)}{v^j(u_{m+2})}, \forall j = 1, \dots, m+1.
\end{equation}
\end{remark}

A k-ad $x=(x_1, \dots, x_k) $ is said to be {\em in general position} if the set of labels of its points includes an ordered subset $ F=\{f_1,\dots,f_{m+2}\}$ such that $\pi_F=([\tilde x_{f_1}],\dots,[\tilde x_{f_{m+2}}])$ is a projective frame. The projective coordinates
$([y_1],\dots,[y_{k-{m-2}}])$ of the remaining landmarks with respect to $\pi_F$, listed in the order they appear in the the k-ad, give a
multivariate axial representation of the projective shape of the k-ad $([\tilde x_{1}],\dots,[\tilde x_{k}]),$ therefore projective shape analysis of k-ads in $P\Sigma^k_m$ is reduced to data analysis on $(\mathbb{R}P^m)^q, q=k-m-2$ (see Patrangenaru and Ellingson (2015)\cite{PaEl:2015})

\subsection{Oriented projective shape and multivariate directional data analysis}

The m dimensional {\em oriented projective space} $\overrightarrow{\mathbb{R}P}^m$ is the quotient $\mathbb R^{m+1}\backslash\{0\}\slash{\sim}$ where $x \sim y$ iff there is a positive scalar $\lambda$ such that $x = \lambda y.$ The equivalence class of $x \in \mathbb R^{m+1}\backslash\{0\}$ is an oriented projective point, labeled $\overrightarrow{[x]}.$ Oriented projective geometry identifies points on rays originating from the null vector in $\R^{m+1}$. Such rays are identified with a unit vectors on the sphere
\[
\SP^m = \{x\in\R^{m+1} : \|x\|=1\},
\]
therefore we may identify $\overrightarrow{\mathbb{R}P}^m$ with $\SP^m.$ Each nonzero vector $x$ defines a direction $x/\|\tilde x\| \in \SP^m$. The $\pm$ sign distinguishes now opposite directions.
An {\em oriented projective frame} in $\overrightarrow{\mathbb{R}P}^m$ is an ordered set of $m + 2$ oriented projective points in
general position $(\overrightarrow{[x_1]},\ldots,\overrightarrow{[x_{m+2}]})$, such that $x_{m+2}$ is a convex linear combination of $x_1,\dots,x_{m+1}.$ An example of oriented projective frame is the {\it standard oriented projective frame} $(\overrightarrow{[e_1]},\dots,\overrightarrow{[e_{m +1}]},\overrightarrow{[e_1 + \ldots + e_{m + 1}]})$ in $\overrightarrow{\mathbb{R}P}^m$.

In OPS analysis one considers oriented k-ads. An oriented k-ad is a k-ad $(\overrightarrow{[x_k]},\dots,\overrightarrow{[x_k]})$ that includes an oriented projective frame.
\begin{definition}
The \emph{oriented projective group} is \(\OPGL(m)\), the set of projective transformations \(g_P\), depending on matrices \(P\in\GLp(m{+}1,\R)\) (positive determinant), acting by \(g_P(\overrightarrow{[x]})=\overrightarrow{[Px]}\).
\end{definition}

Note that $OPGL(m)$ acts simply transitively on oriented projective
frames (see Stolfi(1991)\cite{St:1991}), therefore the action on $\overrightarrow{\mathbb{R}P}^m$ is
determined by its action on a projective frame, thus allowing one to
define the {\em{oriented projective coordinate(s)}} of a point $p \in
\overrightarrow{\mathbb{R}P}^m$ w.r.t. an oriented projective frame  $\pi = (p_1,\dots, p_{m +
2})$ as being given by
\begin{equation} \label{projcoord}
p^{\pi} = \beta^{-1} (p), \end{equation} where $ \beta \in \OPGL(m)$
is the unique projective transformation taking the standard oriented projective frame
to $\pi.$

\begin{remark}\label{r:o-proj-coord} Assume $u, u_1, \dots, u_{m+2}$
are points in $\mathbb{R}^{m},$ such that $\pi = ( \overrightarrow{[\tilde{u}_1]},
\dots, \overrightarrow{[\tilde{u}_{m+2}]})$ is an oriented projective frame. If we consider the
$(m+1) \times (m+1) $ matrix $U_m = [\tilde{u}_1^T, \dots,
\tilde{u}_{m+1}^T],$ the oriented projective coordinates of $ p =\overrightarrow{[\tilde{u}]}$
w.r.t. $\pi$ are given by
\begin{equation} \label{o-projcoord_formula}
 p^{\pi} = \overrightarrow{[ y^1(u), \dots, y^{m+1}(u)]},
\end{equation}
where
\begin{eqnarray} \label{o-v}
v(u) = U_m^{-1} \tilde{u}^T
\end{eqnarray}
and
\begin{equation} \label{o-y}
y^j(u)=\frac{v^j(u)}{v^j(u_{m+2})}, \forall j = 1, \dots, m+1.
\end{equation}
\end{remark}

Fix an oriented projective frame $\pi=(p_{f_1},\dots,p_{f_{m+2}})$ selected from an oriented k-ad and for each label $f \notin F,$ find the corresponding vector $y_f=(y_f^1,\dots,y_f^{m+1})\in \mathbb R^{m+1}$ using Remark \ref{r:o-proj-coord}. Then
$w_f = \frac{y_f}{\|y_{f}\|}\in\SP^m.$

Thus the OPS of the oriented k-ad, is now registered as a point
\[w \in (\SP^m)^q,
\]
where $w=(w_f)_{f\notin F}$. In what follows we work directly with the normalized coordinates $w\in (\SP^m)^q$.

\section{OPS Extrinsic Total Variance}
\label{sec:ops}

We now define an OPS total-variance index in the extrinsic Fr\'echet framework, starting from the directional embedding and then specializing to the planar single-oriented projective coordinate case.

\subsection{Directional embedding and extrinsic Fr\'echet mean}

The basic idea of extrinsic Fr\'echet statistics is to embed the manifold of interest into a Euclidean space, perform Euclidean calculations there, and then project back to the manifold \citep{PaEl:2015}. For OPS we use the inclusion map as  embedding
\[
\jdir:(\SP^m)^q\hookrightarrow(\R^{m+1})^q,
\qquad
\jdir(u_1,\dots,u_q)=(u_1,\dots,u_q).
\]
The ambient space is $\R^{(m+1)q}$ with its usual Euclidean norm.

Let $U=(U^{(1)},\dots,U^{(q)})$ be an $(\SP^m)^q$-valued random variable with distribution $\nu$. The \emph{extrinsic Fr\'echet function} associated with $\jdir$ is
\[
F_{\mathrm{E}}(z)
=
\E\bigl\|\jdir(U)-\jdir(z)\bigr\|^2,
\qquad
z\in(\SP^m)^q.
\]
Any minimizer of $F_{\mathrm{E}}$ is called an \emph{extrinsic Fr\'echet mean} of $U$ with respect to $\jdir$.

In the single oriented projective coordinate case $q=1$ we write $U^{(1)}\in\SP^m$ and
\[
\mu = \E[U^{(1)}]\in\R^{m+1}.
\]
Assuming $\mu\neq0$, the extrinsic mean direction is obtained by projecting $\mu$ back to the sphere:
\[
\mu_{\mathrm{E}} = \frac{\mu}{\|\mu\|}\in\SP^m.
\]
For a sample $\{u_i\}_{i=1}^n\subset\SP^m$ the empirical counterparts are
\[
\bar{u} = \frac{1}{n}\sum_{i=1}^n u_i,
\qquad
\hat\mu_{\mathrm{E}} = \frac{\bar{u}}{\|\bar{u}\|}
\]
provided $\bar{u}\neq0$.

The quantity
\[
R_n = \|\bar{u}\|
\]
is the sample \emph{mean resultant length} from directional statistics and lies in $[0,1]$. It is close to $1$ when the $u_i$ are highly concentrated on the sphere and significantly smaller than $1$ when the $u_i$ are more spread out.

\subsection{Population and sample total-variance indices}

The extrinsic covariance matrix $\Sigma_{\mathrm{E}}$ of $U^{(1)}$ is given in Bhattacharya and Patrangenaru (2005) \cite{BhPa:2005}. Using that closed formula, one can easily show that when q=1, the population OPS extrinsic total-variance is
\[
\tSigmaOPS = 2(1-R)
= 2\bigl(1-\|\E[U^{(1)}]\|\bigr).
\]

The OPS extrinsic total variance vanishes if and only if $U^{(1)}$ is almost surely constant, and increases as the distribution of $U^{(1)}$ spreads out on the sphere.

Given an OPS random sample of oriented (m+3)-ads in their spherical representation  $\{u_i\}_{i=1}^n\subset\SP^m$, the sample OPS extrinsic total variance is
\begin{equation}
  \tSOPS = 2(1-R_n),
  \qquad
  R_n = \|\bar{u}\|.
  \label{eq:tS-ops}
\end{equation}
In the Sope Creek application we have $m=2$ and $q=1$, so \eqref{eq:tS-ops} completely describes the OPS total variance in terms of the mean of the three-dimensional unit vectors $\{u_i\}$.


\subsection{VW total-variance and VW mean axis}
For a comparison we review the extrinsic total-variance index in the PS model based on the Veronese--Whitney embedding.
For $[x]\in\RP^m$ with $x\in\R^{m+1}\setminus\{0\}$, the Veronese--Whitney embedding is defined by
\[
\jVW([x]) = \frac{xx^\top}{\|x\|^2}
\in\mathrm{Sym}_{m+1},
\]
where $\mathrm{Sym}_{m+1}$ denotes the real symmetric $(m{+}1)\times(m{+}1)$ matrices. This mapping is invariant under nonzero rescaling of $x$, and therefore is well defined on projective space.

Let $X$ be an $\RP^m$-valued random variable. In the single-coordinate case we consider the extrinsic mean matrix
\[
K = \E[\jVW(X)].
\]
Under a nonfocality condition, $K$ has a unique largest eigenvalue $\lambda_1$ with eigenvector $v_1$, and the corresponding axis $\pm v_1$ represents the VW extrinsic projective shape; see \citet{PaEl:2015}.

In the case q=1, Fisher et al (1996)\cite{FiHaJiWo:1996} showed that VW total-variance for PS is given by
\[
\tSigmaPS = 2(1-\lambda_1).
\]
Given a sample $\{X_i\}_{i=1}^n$ with homogeneous representatives $\{z_i\}_{i=1}^n\subset\R^{m+1}\setminus\{0\}$, we form
\[
J_n = \frac{1}{n}\sum_{i=1}^n \frac{z_i z_i^\top}{\|z_i\|^2},
\]
let $\lambda_1^{(n)}$ be the largest eigenvalue of $J_n$, and define the sample index
\begin{equation}
  \tSPS = 2\bigl(1-\lambda_1^{(n)}\bigr).
  \label{eq:tS-ps}
\end{equation}
The hypothesis $H_0:\tSigmaPS=0$ corresponds to the remaining landmark being projectively constant after frame normalization.

\section{OPS Coplanarity Test and Asymptotic Standard Error}
\label{sec:test}

We now derive a delta-method based standard error for $\tSOPS$ and formulate a large-sample test for coplanarity in the OPS model.

\subsection{Sample covariance and delta-method variance}

Let $\{u_i\}_{i=1}^n\subset\SP^m$ be OPS coordinates of a remaining landmark after oriented frame normalization. Define
\[
\bar{u} = \frac{1}{n}\sum_{i=1}^n u_i,
\qquad
R_n = \|\bar{u}\|,
\qquad
\tSOPS = 2(1-R_n).
\]
We use the sample covariance matrix with $1/n$ normalization,
\[
S_n
=
\frac{1}{n}\sum_{i=1}^n (u_i-\bar{u})(u_i-\bar{u})^\top.
\]

Consider the smooth map $g:\R^{m+1}\to\R$ given by
\[
g(x) = 2(1-\|x\|).
\]
Its gradient at $\bar{u}$ is
\[
\nabla g(\bar{u}) = -\,2\,\frac{\bar{u}}{\|\bar{u}\|}.
\]
Under standard moment and nondegeneracy assumptions, the central limit theorem and the delta-method yield
\[
\sqrt{n}\bigl(\tSOPS - \tSigmaOPS\bigr)
\Rightarrow
\mathcal{N}\bigl(0,\sigma^2_{\OPS}\bigr),
\]
with asymptotic variance
\[
\sigma^2_{\OPS}
=
4\,\frac{\mu^\top\Sigma_{\mathrm{E}}\mu}{\|\mu\|^2},
\]
and empirical estimator
\begin{equation}
  \widehat{\Var}(\tSOPS)
  \approx
  \frac{1}{n}\,\nabla g(\bar{u})^\top S_n \nabla g(\bar{u})
  =
  \frac{4}{n}\,\frac{\bar{u}^\top S_n \bar{u}}{\|\bar{u}\|^2}.
  \label{eq:var-ops}
\end{equation}
We write
\begin{equation}
  \widehat{\mathrm{SE}}_{\OPS}
  =
  \sqrt{\widehat{\Var}(\tSOPS)}
  =
  \frac{2}{\sqrt{n}}\,
  \frac{\sqrt{\bar{u}^\top S_n \bar{u}}}{\|\bar{u}\|}.
  \label{eq:se-ops}
\end{equation}
This is the delta-method (extrinsic CLT-based) standard error used later in our numerical summaries.

We test coplanarity via the one-sided hypothesis
\begin{equation}
  H_0:\ \tSigmaOPS = 0
  \qquad\text{vs.}\qquad
  H_A:\ \tSigmaOPS > 0.
  \label{eq:ops-hypothesis}
\end{equation}
For a nominal level $\alpha\in(0,1)$, let $z_{1-\alpha/2}$ denote the standard normal quantile. A delta-method confidence interval for the population index $\tSigmaOPS$ is
\begin{equation}
  \tSOPS
  \pm
  z_{1-\alpha/2}\,\widehat{\mathrm{SE}}_{\OPS}.
  \label{eq:ci-ops}
\end{equation}
Because the coplanarity hypothesis is one-sided, we test \eqref{eq:ops-hypothesis} by examining the lower endpoint of \eqref{eq:ci-ops}. At level $\alpha$ we reject $H_0$ if this lower endpoint is strictly positive.

Equivalently, we form the $Z$-statistic
\[
Z_{\OPS}
=
\frac{\tSOPS}{\widehat{\mathrm{SE}}_{\OPS}},
\]
which is approximately standard normal under $H_0$ when dispersion is small. The associated one-sided $p$-value is $1-\Phi(Z_{\OPS})$.

As a complementary calibration we consider the scaled statistic
\[
T_{\OPS} = n\tSOPS
\]
and compare it to a $\chi^2_2$ reference when $m=2$. In this planar single-coordinate case the OPS manifold has a two-dimensional tangent space at the extrinsic mean direction, which motivates the use of a chi-square distribution with two degrees of freedom as an asymptotic approximation for $T_{\OPS}$.

\subsection{Leave-out-outliers influence and reduced samples}

In moderate size samples it is natural to examine the influence of individual scenes on $\tSOPS$. For $i=1,\dots,n$ let $\tSOPS^{(-i)}$ be the OPS total-variance index computed after deleting scene $i$, and let $\widehat{\mathrm{SE}}_{\OPS}^{(-i)}$ be the corresponding standard error. The associated $Z$-statistics or the lower endpoints of the intervals \eqref{eq:ci-ops} can be used to rank scenes by influence.

In the Sope Creek analysis we use a simple greedy scheme at a reference level $\alpha_{\mathrm{ref}}=0.05$: starting from the full sample, we remove the scene whose deletion yields the largest increase in the lower endpoint of the OPS interval at level $\alpha_{\mathrm{ref}}$. The process stops when this lower endpoint first becomes nonpositive. The remaining scenes form a reduced sample. We report both full-sample and reduced-sample statistics for OPS and PS, with the reduced sample defined by OPS.


\subsection{Data and oriented projective frame}
We now apply the OPS and PS total-variance indices to the Sope Creek stone, using the oriented projective frame $F=\{1,2,4,3\}$.

The Sope Creek dataset consists of $n=41$ photographs of a stone in a stream at Sope Creek, Marietta, GA \citep{Patrangenaru2001}. Each image contains $k=5$ landmarks placed at distinct, visually recognizable points on the stone's contour.
 \begin{figure}[H]
  \centering
  \includegraphics[scale=2.2]{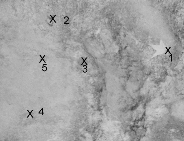}
  \caption{Landmarks for the Sope Creek stone: frame landmarks \(\{1,2,4,3\}\) and remaining landmark \(5\).
  Adapted from Figure 1.10 in \citet{PaEl:2015}, originally due to \citet{Patrangenaru2001}.}
  \label{fig:sopecreek-frame}
\end{figure}

Following the projective shape literature, the aim is to assess whether the stone surface is projectively flat (coplanar) under central projection.

Throughout most of the analysis we use the oriented frame
\[
F = \{1,2,4,3\},
\qquad
R = \{5\},
\]
so that landmarks $1,2,4,3$ form a projective frame and landmark $5$ serves as the single remaining coordinate. For each scene $i$ we form the $3\times4$ matrix $P_i$ from the homogeneous coordinates of the frame landmarks, compute a homography $H_i$ mapping $P_i$ to a canonical oriented frame $Q=[e_1,e_2,e_3,\mathbf{1}]$ (with $\mathbf{1}=(1,1,1)^\top$), and choose the representative $H_i$ with $\det(H_i)>0$. The oriented projective coordinate of landmark $5$ in scene $i$ is then
\[
u_i = \frac{H_i\tilde x_{i,5}}{\|H_i\tilde x_{i,5}\|}\in\SP^2,
\qquad
i=1,\dots,41.
\]
These unit vectors are the input to OPS analysis under the selected OP frame.

\subsection{OPS geometry under the OP frame}

Under the frame $F=\{1,2,4,3\}$ the sample mean direction and mean resultant length are
\[
\bar{u}
\approx
(0.0073,\,-0.6720,\,-0.6082)^\top,
\qquad
R_n \approx 0.9064.
\]
Thus the oriented coordinates are tightly concentrated on a spherical cap of $\SP^2$ pointing roughly toward the negative $yz$-hemisphere.

\begin{figure}[H]
  \centering

  \begin{subfigure}[t]{0.6\textwidth}
    \centering
    \includegraphics[width=\textwidth]{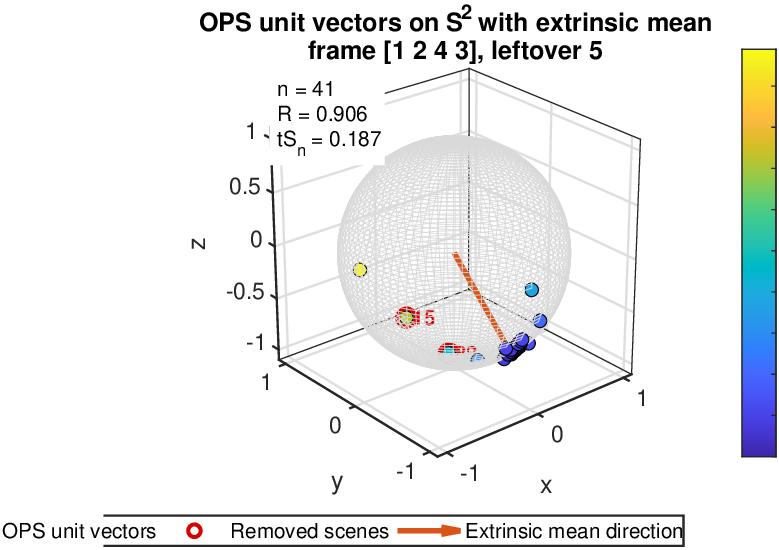}
    \caption{Full-sphere view under frame $\{1,2,4,3\}$ with remaining landmark $5$.
    The arrow shows the extrinsic mean direction and removed scenes are circled.}
    \label{fig:ops-sphere-full}
  \end{subfigure}

  \vspace{0.8em}

  \begin{subfigure}[t]{0.6\textwidth}
    \centering
    \includegraphics[width=\textwidth]{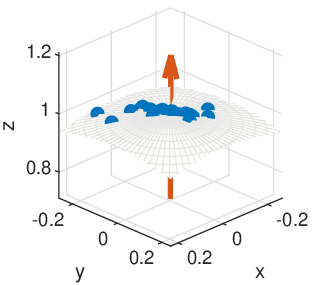}
    \caption{Zoomed view near the extrinsic mean under the same frame,
    making the influential scenes easier to see.}
    \label{fig:ops-sphere-zoom}
  \end{subfigure}

  \caption{OPS unit vectors on $\mathbb{S}^2$ for the Sope Creek stone under
  frame $\{1,2,4,3\}$ with remaining landmark $5$: (a) full-sphere view;
  (b) zoomed view near the extrinsic mean.}
  \label{fig:ops-sphere}
\end{figure}

A natural method of visualization is to plot the OPS unit vectors $\{u_i\}$ on the sphere, with the extrinsic mean direction indicated by an arrow and influential scenes highlighted. In Figure~\ref{fig:ops-sphere} we show a full-sphere view together with a zoomed view that emphasizes the data cloud and the removed scenes under the primary frame choice.

The angular distances between $u_i$ and the extrinsic mean
$\hat\mu_{\mathrm{E}}$ provide a one-dimensional summary of dispersion.
Histograms of these angles, for the full and reduced samples, give a direct
sense of how much the data tighten after removing influential scenes.

\begin{figure}[t]
  \centering
  \includegraphics[width=0.75\textwidth]{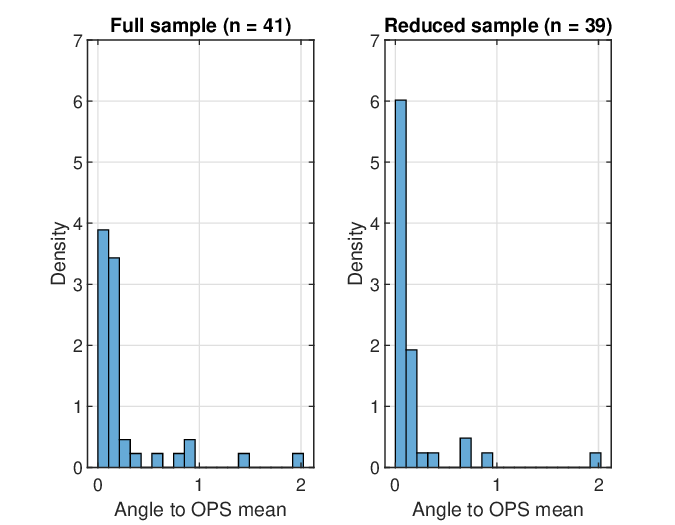}
  \caption{Angular distances between OPS unit vectors and the extrinsic mean
  under $F=\{1,2,4,3\}$ for the full sample (left) and the OPS-based reduced
  sample (right).}
  \label{fig:angle-hist-F}
\end{figure}

\subsection{Full-sample OPS under the OP frame}

Using \eqref{eq:tS-ops}, the full-sample OPS total-variance index under $F$
with $n=41$ is
\[
\tSOPS \approx 0.1871.
\]
The scaled statistic is
\[
T_{\OPS}
=
n\tSOPS
\approx
41\times0.1871
=
7.6731,
\]
which yields a chi-square $p$-value
\[
p_{\chi^2,\OPS} \approx 0.0216
\]
under a $\chi^2_2$ reference.

From the sample covariance matrix $S_n$ and \eqref{eq:se-ops} we obtain
\[
\widehat{\mathrm{SE}}_{\OPS} \approx 0.0812.
\]
The corresponding $95\%$ delta-method confidence interval is
\[
0.1871 \pm 1.96\times0.0812 \;=\; [0.028,\,0.346].
\]
The lower endpoint is strictly positive, so both the chi-square calibration
and the normal approximation reject $H_0:\tSigmaOPS=0$ at the $5\%$ level in
the full sample under the primary frame.

The corresponding delta-method standard error is about $0.076$, leading to a
$95\%$ interval with a positive lower endpoint. Thus, under the primary
frame $F$, the OPS detects a small but statistically significant
deviation from perfect coplanarity in the full sample.

\subsection{OPS-based reduced sample under the OP frame}

Applying the OPS leave-one-out diagnostic of Section~\ref{sec:test} with
reference level $\alpha_{\mathrm{ref}}=0.05$ under the frame $F$ identifies a
pair of scenes with the largest influence on the lower endpoint of the OPS
interval. Removing these two scenes yields a reduced sample of size
$n_{\mathrm{red}}=39$.

On this reduced sample the OPS statistics under $F$ become
\[
R_n^{(\mathrm{red})} \approx 0.9352,
\qquad
\tSOPS^{(\mathrm{red})} \approx 0.1297,
\]
and
\[
T_{\OPS}^{(\mathrm{red})}
=
39\times 0.1297
\approx
5.0581,
\]
with chi-square $p$-value
\[
p_{\chi^2,\OPS}^{(\mathrm{red})} \approx 0.0797.
\]
The reduced-sample standard error
$\widehat{\mathrm{SE}}_{\OPS}^{(\mathrm{red})}\approx0.0758$ yields a $95\%$
interval
\[
0.1297 \pm 1.96\times0.0758 \;=\; [-0.019,\,0.278].
\]
The lower endpoint is slightly negative, so at the $5\%$ level the
large sample one-sided OPS delta-method test no longer rejects $H_0:\tSigmaOPS=0$ for the
reduced sample under $F$. In this sense, after removing two influential
scenes the OPS analysis under the primary frame is compatible with exact
coplanarity, and with the results for concentrated PS coplanarity of Sope Creek stone data analysis in Patrangenaru(2001)\cite{Patrangenaru2001}, where on a reduced sample ($n=37$) it was reported  a total-variance estimate $\hat t_{n,I}\approx 0.58$ with standard error $\mathrm{SE}(\hat t_{n,I})\approx 2.63$, using inhomogeneous local coordinates, concluding that the coplanarity (flatness) hypothesis could not be rejected at $\alpha=0.05$.

Our extrinsic OPS analysis, using a related frame ordering and an OPS-based removing two outliers to $n=39$ under $F=\{1,2,4,3\}$ , reaches a qualitatively similar conclusion for the reduced samples: the oriented surface is compatible with coplanarity at the $5\%$ level. At the same time, working on $\mathbb{S}^2$ provides a transparent ``almost flat" geometry: the OPS index has the closed form $\tSOPS=2(1-R_n)$ with an interpretable delta-method standard error.

\section{Acknowledgement and Discussion}
\label{sec:discussion}

The research in this manuscript was supported by the National Science Foundation under Grants DMS - 2311058 (Paige), DMS - 2311059 (Patrangenaru).

The OPS framework extends classical PS results by differentiating real scenes from their virtual images by means of orientation-preserving projective transformations. Within this framework, the OPS extrinsic total-variance index
\[
\tSOPS = 2(1-R_n),
\qquad
R_n = \biggl\|\frac{1}{n}\sum_{i=1}^n u_i\biggr\|,
\]
provides a simple scalar measure of dispersion of oriented projective coordinates around their extrinsic mean direction. It is expressed directly in terms of the mean resultant length on the sphere and fits naturally into the extrinsic Fr\'echet theory of \citet{PaEl:2015}.

The Sope Creek analysis illustrates how  OPS extrinsic total variance can be used in practice when a large sample of images of a 3D scene are available.

From a broader perspective, the OPS extrinsic total-variance index complements the OPS mean-based methods developed in \citet{ChPaPa:2022}. Their work emphasizes OPS spaces and tests for differences in oriented mean shape, while the present contribution focuses on the magnitude of dispersion around a single OPS mean direction.

Note that for higher-dimensional scenes (such as $m=3$ ), the OPS total-variance approach becomes difficult, due to the inability to observe the  $m+2$ landmark as a convex combination of the previous ones. Finally, OPS total variance can be used in face recognition.

Tha table with corresponding landmark coordinates on Sope Creek stone data, from a sample of 41 images captured with a classical pinhole camera, and later digitized, can be found in Chapter 1 of Patrangenaru and Ellingson (2015)\cite{PaEl:2015}.

All computations described in this paper (including oriented projective alignment, construction of unit vectors on $\mathbb{S}^2$, Veronese--Whitney covariance estimation) were carried out in \textsc{MATLAB} R2024b.

The scripts that generate the figures, tables, and numerical summaries reported in this paper are available from the authors upon request.


\newpage
\bibliographystyle{plainnat}

\end{document}